\documentclass[conference,a4paper,article]{IEEEtran}
\usepackage{amsmath,amssymb,epsfig,color,flushend,epstopdf,algorithm}
\usepackage{algorithm,epstopdf}
\usepackage[noend]{algpseudocode}
\makeatletter
\def\BState{\State\hskip-\ALG@thistlm}
\makeatother

\begin{document}
\title{Two-Way Relay Beamforming Optimization for Full-Duplex SWIPT Systems}

\author{\IEEEauthorblockN{Alexander A. Okandeji, Muhammad R. A. Khandaker, and Kai-Kit Wong}
\IEEEauthorblockA{Department of Electronic and Electrical Engineering\\
University College London\\
Gower Street, London, WC1E 7JE, United Kingdom\\
e-mail: $\{\rm alexander.okandeji.13, m.khandaker, kai\text{-}kit.wong\}@ucl.ac.uk$}\\
\textit{(Invited Paper)}
}
\maketitle

\begin{abstract}
In this paper, we investigate the problem of two-relay beamforming optimization to maximize the achievable sum-rate of a simultaneous wireless information and power transfer (SWIPT) system with a full-duplex (FD) multiple-input multiple-output (MIMO) amplify-and-forward (AF) relay. In particular, we address the optimal joint design of the receiver power splitting (PS) ratio and the beamforming matrix at the relay given the channel state information (CSI). Our contribution is an iterative algorithm and one-dimensional (1-D) search to achieve the joint optimization. Simulation results are provided to demonstrate the effectiveness of the proposed algorithm.
 \end{abstract}
 
\section{Introduction}\label{sec_intro}
Conventionally, wireless communication nodes operate in half duplex (HD) mode under which they transmit and receive signals over orthogonal frequency or time resources. Recent advances, nevertheless, suggest that full duplex (FD) communications that allows simultaneous transmission and reception of signal over the same radio channel be possible \cite{exp,method_broad}. 

In addition to the immediate benefit of essentially doubling the bandwidth, full duplex communications also find applications in simultaneous wireless information and power transfer (SWIPT). Much interest has turned to full-duplex relaying in which information is sent from a source node to a destination node through an intermediate relaying node which is powered by means of wireless energy harvesting. In the literature, the studies on relay aided SWIPT largely considered HD relaying and adopted a time-switched relaying (TSR) approach \cite{RF}--\cite{Na}. 

Authors in \cite{ruhuul} considered SWIPT in MISO multicasting systems, in \cite{RF2} considered SWIPT in MISO broadcasting systems, and in \cite{ruhul_kk, ruhuul2} MISO secrecy systems, where the joint transmit beamforming and receive power splitting problem for minimising the transmit power of the Base station (BS) subject to signal-to-noise ratio (SNR) and energy harvesting constraints at the receiver was investigated. 

In contrast to the existing results, this paper studies the joint optimization of the two-way beamforming matrix for SWIPT in a multiple-input multiple-output (MIMO) amplify-and-forward (AF) full-duplex relay system employing a power splitter (PS), where the sum rate is maximized subject to the energy harvesting and total power constraints.

{\em Notations}---We use ${\bf X}\in\mathbb{C}^{M \times N}$ to represent a complex matrix with dimension of $M \times N$. Also, we use $(\cdot)^{\dagger}$ to denote the conjugate transpose, while $\mathrm{trace}(\cdot)$ is the trace operation, and $\|\cdot\|$ denotes the Frobenius norm. In addition, $| \cdot|$ returns the absolute value of a scalar, and ${\bf X}\succeq {\bf 0}$ denotes that the Hermitian matrix ${\bf X}$ is positive semidefinite. The expectation operator is denoted by $\mathbb{E}\{\cdot\}.$ We define $\Pi_\mathbf{X} = \mathbf{X}(\mathbf{X}^{\dagger}\mathbf{X})^{-1}\mathbf{X}^{\dagger}$ as the orthogonal projection onto the column space of $\mathbf{X}$; and $\Pi^{\perp}_{\mathbf{X}} = \mathbf{I} - \Pi_\mathbf{X}$ as the orthogonal projection onto the orthogonal complement of the column space of $\mathbf{X}.$

\begin{figure}
\begin{center}
\includegraphics[width=8cm]{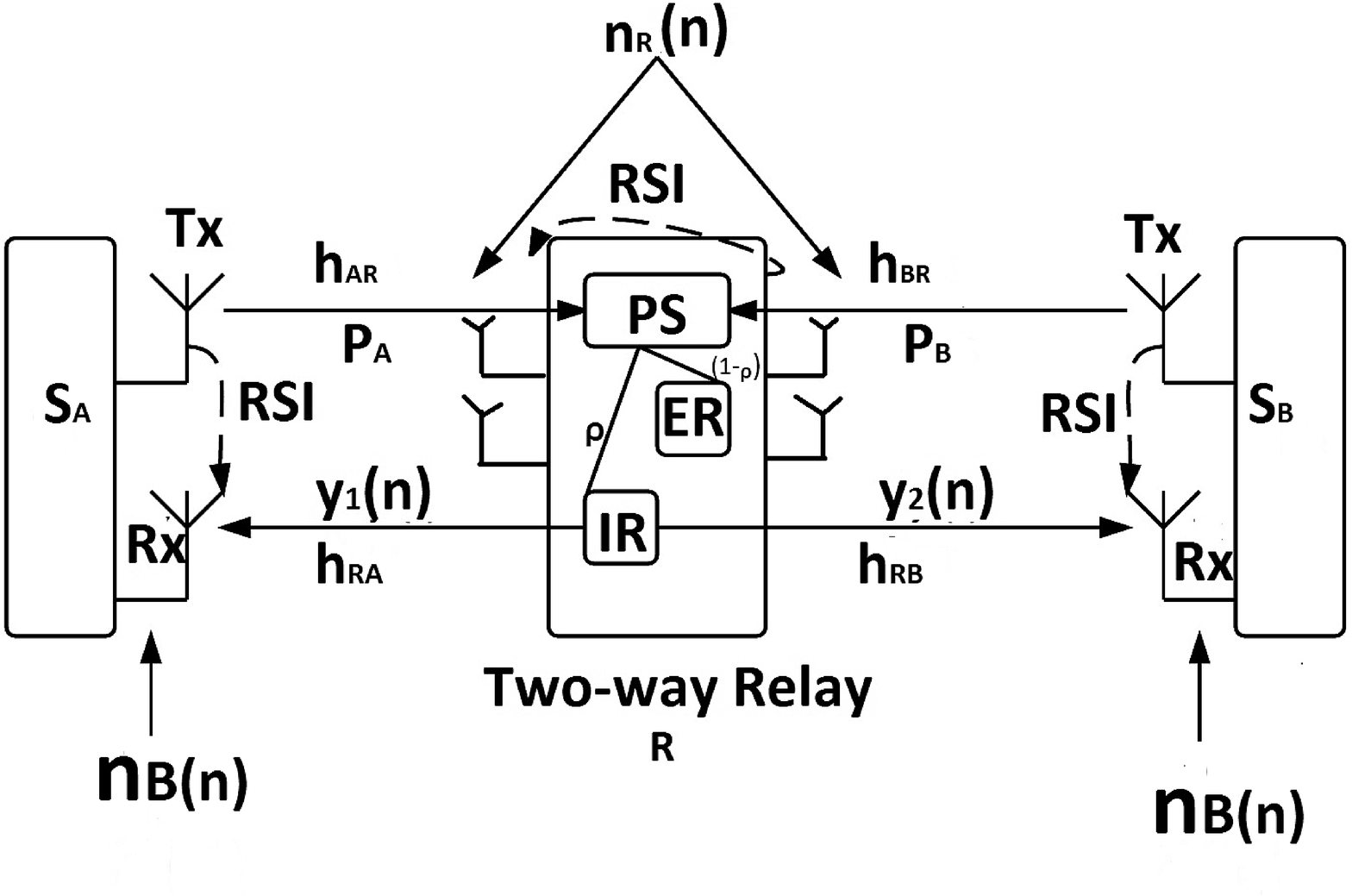}
\caption{The model of the two-way full-duplex SWIPT system.}\label{alex}
\end{center}
\end{figure}

\section{System Model}
Let us consider SWIPT in a three-node MIMO relay network consisting of two sources ${\sf S_A}$ and ${\sf S_B}$ wanting to exchange information with the aid of an AF relay ${\sf R}$, as shown in Fig.~\ref{alex}. In our model, all the nodes are assumed to operate in FD mode, and we also assume that there is no direct link between ${\sf S_A}$ and ${\sf S_B}$ so communication between them must be done via ${\sf R}$. Both ${\sf S_A}$ and ${\sf S_B}$ transmit their messages simultaneously to ${\sf R}$ with transmit power ${P}_A$ and ${P}_B$, respectively. 

In the broadcast phase, the relay ${\sf R}$ employs linear processing with an amplification matrix $\mathbf{W}$ to process the received signal and broadcasts the processed signal to the nodes with the harvested power $Q$. We assume that each source node is equipped with a pair of transmitter-receiver antennas for signal transmission and reception respectively.  We use M$_T$ and M$_R$ to denote the number of transmit and receive antennas at ${\sf R}$, respectively. We use $\mathbf{h}_{XR} \in \mathbb{C}^{M_R \times 1}$ and $\mathbf{h}_{RX}\in \mathbb{C}^{M_T \times 1} $ to, respectively, denote the directional channel vectors between the source node $X$'s $(\in A,B)$ transmit antenna to ${\sf R}$'s receive antennas, and that between the relay's transmit antenna(s) to source node $X$'s receive antenna. The concurrent transmission and reception of signals at the nodes produces self-interference (SI) which inhibits the performance of a full duplex system. We consider using existing SI cancellation mechanisms in the literature to mitigate the SI (e.g., antenna isolation, analog and digital cancellation, and etc.) \cite{As}. 

Due to imperfect channel estimation, however, the SI cannot be cancelled completely \cite{single_channel}. We therefore denote ${h_{AA}},$ ${h_{BB}}$ and $\mathbf{H}_{RR} \in \mathbb{C}^{M_R \times M_T}$ as the SI channels at the corresponding nodes. For simplicity, we model the residual SI (RSI) channel as a Gaussian distribution random variable with zero mean and variance $\sigma^2_{X}$, for $X\in \{A,B,R\}$ \cite{single_channel}. We further assume that the relay is equipped with a PS device which splits the received signal power at the relay for energy harvesting, amplification and forwarding of the received signal. In particular, the received signal at the relay is split such that a $\rho\in (0,1)$ portion of the received signal power at the relay is fed to the information receiver (IR) and the remaining $(1-\rho)$ portion of the power to the energy receiver (ER) at the relay. 

When the source nodes transmit their signals to the relay, the AF relay employs a short delay to perform linear processing. It is assumed that the processing delay at the relay is given by a $\tau$--symbol duration, which denotes the processing time required to implement the full duplex operation \cite{tau}. $\tau$  typically takes integer values. We assume that the delay is short enough compared to a time slot which has a large number of data symbols, and thus its effect on the achievable rate is negligible. At time instant $n,$ the received signal $\mathbf{y}_r[n]$ and the transmit signal $\mathbf{x}_R[n]$ at the relay can be, respectively, written as
\begin{align}
\mathbf{y}_r[n] &= \mathbf{h}_{AR}{s_A}[n] + \mathbf{h}_{BR}{s_B}[n]+ \mathbf{H}_{RR} \mathbf{x}_R[n] + \mathbf{n}_R[n],\label{y1}\\
\mathbf{x}_R[n] &=   \mathbf{W y}^{\rm IR}_r(n-\tau),\label{X_r}
\end{align}
where $\mathbf{y}_r^{\rm IR}[n]$ is the signal split to the IR at ${\sf R}$ given by 
\begin{multline}\label{y21}
 \mathbf{y}^{\rm IR}_r[n]= \rho(\mathbf{h}_{AR}{s_A}[n] +  \mathbf{h}_{BR}{s_B}[n]+\mathbf{H}_{RR} \mathbf{x}_R[n]\\
 + \mathbf{n}_R[n])  +n_p[n],
\end{multline}
where $n_p\sim{\cal CN}$ $(0, \sigma_p^2)$ is the additional processing noise at the IR. Using (\ref{X_r}) and (\ref{y21}) recursively, the overall relay output can be rewritten as
\begin{multline}
\mathbf{x}_R[n]=\mathbf{W} (\rho(\mathbf{h}_{AR}{s_A}[n-\tau] + \mathbf{h}_{BR}{s_B}[n-\tau]\\
+\mathbf{H}_{RR} \mathbf{x}_R[n-\tau] + \mathbf{n}_R[n -\tau]) + n_p[n-\tau]).
\end{multline}
The capacity of a relay network with delay depends only on the relative path delays from the sender to the receiver and not on absolute delays \cite{delay}. Thus, the relay output is given as
\begin{eqnarray}
\mathbf{x}_R[n]
\!\!\!& = &\!\!\! \mathbf{W} \sum_{j=0}^{\infty} (\mathbf{H}_{RR} \mathbf{W})^j  [\rho(\mathbf{h}_{AR}{s_A}[n-j\tau-\tau] \nonumber\\
\!\!\!& + &\!\!\!  \mathbf{h}_{BR}{S_B}[n-j\tau-\tau] + \mathbf{n}_R[n-j\tau-\tau]) \nonumber\\
\!\!\!& + &\!\!\! n_p[n-j\tau-\tau]], \label{xr_new}
\end{eqnarray}
where $j$ denotes the index of the delayed symbols.

To simplify the signal model and to keep the optimization problem  more tractable, we add the zero forcing (ZF) solution constraints such that the optimization of $\mathbf{W}$ chooses to null out the RSI from the relay output to the relay input \cite{joint}. To realize this, it is easy to check from (\ref{xr_new}) that the following condition is sufficient \cite{joint}:
\begin{equation}\label{approx}
\mathbf{W}\mathbf{H}_{RR}\mathbf{W} ={\bf 0}.
\end{equation}
Consequently, (\ref{xr_new}) becomes 
\begin{eqnarray}
\mathbf{x}_R[n]\!\!\!& = &\!\!\!\mathbf{W} (  \rho (\mathbf{h}_{AR}{s_A}[n-\tau] + \mathbf{h}_{BR}{s_B}[n-\tau] \nonumber\\
\!\!\!& + &\!\!\! \mathbf{n}_R[n-\tau]) +  n_p[n-\tau]), \label{xr_new1}
\end{eqnarray}
with the covariance matrix 
\begin{eqnarray}
\mathbb{E} \{\mathbf{x}_R\mathbf{x^{\dagger}}_R\}\!\!\!& = &\!\!\! \rho P_A \mathbf{W}\mathbf{h}_{AR} \mathbf{h^{\dagger}}_{AR} \mathbf{W^{\dagger}} + \rho P_B \mathbf{W}\mathbf{h}_{BR} \mathbf{h^{\dagger}}_{BR} \mathbf{W^{\dagger}}\nonumber\\
\!\!\!& + &\!\!\! \rho \mathbf{W} \mathbf{W^{\dagger}} + \mathbf{W} \mathbf{W^{\dagger}}.\label{Sigma}
\end{eqnarray}
Thus the relay output power can be written as 
\begin{multline}\label{Pr_new}
p_R=\mathrm{trace}(\mathbb{E} \{\mathbf{x}_R\mathbf{x^{\dagger}}_R\}) =   \rho [P_A\| \mathbf{W} \mathbf{h}_{AR}\|^2\\
+P_B\| \mathbf{W} \mathbf{h}_{BR}\|^2 
  + \mathrm{trace}(\mathbf{W}\mathbf{W}^\dagger)] + \mathrm{trace} (\mathbf{W}\mathbf{W}^\dagger).
\end{multline}
The received signal at ${\sf S_A}$ is given by
\begin{eqnarray}
y_{s_A}[n] \!\!\!& = &\!\!\!\mathbf{h^{\dagger}}_{RA} \mathbf{x}_R[n] + h_{AA}s_A[n] + n_A[n]\nonumber\\
\!\!\!& = &\!\!\! \rho( \mathbf{h^{\dagger}}_{RA} \mathbf{W} \mathbf{h}_{AR} s_A [n-\tau] \nonumber\\
\!\!\!& + &\!\!\!  \mathbf{h^{\dagger}}_{RA} \mathbf{W} \mathbf{h}_{BR} s_B [n-\tau] + \mathbf{h^{\dagger}}_{RA}\mathbf{Wn}_R[n])\nonumber\\
\!\!\!& + &\!\!\! \mathbf{h^{\dagger}}_{RA} \mathbf{W}n_p[n] + h_{AA}s_A[n]+ n_A[n].\label{y^{SA}}
\end{eqnarray}
After cancelling its own signal $s_A[n-\tau],$ it becomes
\begin{eqnarray}
y_{s_A}[n] \!\!\!& = &\!\!\! \rho( \mathbf{h^{\dagger}}_{RA} \mathbf{W} \mathbf{h}_{BR} s_B [n-\tau] + \mathbf{h^{\dagger}}_{RA}\mathbf{Wn}_R[n])\nonumber\\
\!\!\!& + &\!\!\! \mathbf{h^{\dagger}}_{RA} \mathbf{W}n_p[n] + h_{AA}s_A[n]+ n_A[n].\label{y^{SA1}}
\end{eqnarray}
The received signal-to-interference-plus-noise ratio (SINR) at node $A$, denoted as $\gamma_A,$ can be expressed as 
\begin{equation}
\gamma_A = \frac{\rho^2 P_B |\mathbf{h^{\dagger}}_{RA} \mathbf{W} \mathbf{h}_{BR}|^2}{\rho^2 \|\mathbf{h^{\dagger}}_{RA} \mathbf{W}\|^2 + \|\mathbf{h^{\dagger}}_{RA} \mathbf{W}\|^2 + P_A| {h_{AA}}|^2 + 1}.
\end{equation}
Similarly, the received SINR $\gamma_B$ at node $B$ can be written as
\begin{equation}
\gamma_B = \frac{\rho^2 P_A |\mathbf{h^{\dagger}}_{RB} \mathbf{W} \mathbf{h}_{AR}|^2}{\rho^2 \|\mathbf{h^{\dagger}}_{RB} \mathbf{W}\|^2 + \|\mathbf{h^{\dagger}}_{RB} \mathbf{W}\|^2 + P_B| {h_{BB}}|^2 + 1}.
\end{equation}
The achievable rates are then given by $R_A = \log_2(1 + \gamma_A)$ and $R_B = \log_2(1+\gamma_B),$ at nodes $A$ and $B$, respectively.

The signal split to the ER at ${\sf R}$ is given as
\begin{equation}\label{Q1}
y^{\rm ER} = \beta(1-\rho)(\mathbf{h}_{AR}{s_A}[n] + \mathbf{h}_{BR}{s_B}[n]+ \mathbf{H}_{RR} \mathbf{x}_R[n] + \mathbf{n}_R[n]),
\end{equation}
where $\beta$ denotes the energy conversion efficiency of the ER at the relay which accounts for the loss in energy transducer for converting the harvested energy to electrical energy to be stored. In this paper, for simplicity, we assume $\beta = 1$. Thus, the harvested energy at the relay is given by
\begin{equation}
Q = (1-\rho)(|\mathbf{h}_{AR}|^2{P_A} + |\mathbf{h}_{BR}|^2{P_B}+ \mathrm{\bar{E}}] + \delta_R),\label{Q1a}
\end{equation}
where $\mathrm{\bar{E}} = \mathbb{E} \{\mathbf{x}_R\mathbf{x^{\dagger}}_R\}$ and $\delta_R$ is the additive white Gaussian noise (AWGN) with zero mean and unit variance at the relay.

Note that the conventional HD relay communication system requires two phases for ${\sf S_A}$ and ${\sf S_B}$ to exchange information \cite{study}. FD relay systems on the other hand reduce the whole operation to only one phase, hence increasing the spectrum efficiency. 
For simplicity, we assume that the transmit power at the source nodes are intelligently selected by the sources. Therefore, in this work, we do not consider optimization at the source nodes. To ensure a continuous information transfer between the two sources, the harvested energy at the relay should be above a given threshold so that a useful level of harvested energy is reached. As a result, we formulate the joint relay beamforming and receive PS ratio ($\rho$) optimization problem as a maximization problem of the sum rate. Mathematically, this problem is formulated as  
\begin{eqnarray}
\max_{{\mathbf{W},\rho \in(0,1)}}  \!\!\!& &\!\!\! R_A + R_B \nonumber\\ {\rm s.t.} \!\!\!& &\!\!\!   Q\geq \bar{Q} \nonumber\\
\!\!\!& &\!\!\!   p_R \leq P_R, \label{Problem}
\end{eqnarray}
where $P_R$ is the maximum transmit power at the relay and $\bar{Q}$ is the minimum amount of harvested energy required to maintain the relay's operation.
  

\section{Proposed Solution}\label{max_sum_rate}
In this section, our aim is to maximize the sum-rate of the proposed FD MIMO two-way AF-relaying channel. Considering the fact that each source only transmits a single data stream and the network coding principle encourages mixing rather than separating the data streams from the two sources, we decompose $\mathbf{W}$ as $\mathbf{W} = \mathbf{w}_t \mathbf{w}_r^{\dagger}$,  where $\mathbf{w}_t$ is the transmit beamforming vector and $\mathbf{w}_r$ denotes the receive beamforming vector at the relay. Then the ZF condition is simplified to $(\mathbf{w}_r^{\dagger}\mathbf{H}_{RR}\mathbf{w}_t)\mathbf{W}={\bf 0}$ or equivalently $\mathbf{w}_r^{\dagger}\mathbf{H}_{RR}\mathbf{w}_t=0$ because in general $\mathbf{W}\neq0$ \cite{joint}. We further assume without loss of optimality that $\|\mathbf{w}_r\| = 1$. Therefore, the optimization problem in (\ref{Problem}) can be rewritten as (\ref{y6}) (see top of next page)
\begin{figure*}[!t]
\normalsize
\begin{subequations}\label{y6}
\begin{align}
 \max_{{\mathbf{w}_r,\mathbf{w}_t  \rho \in(0,1)}} &~~ 
  \log_2\left( 1+ \frac{\rho^2 P_B C_{rB}|\mathbf{h^{\dagger}}_{RA} \mathbf{w}_t|^2 }{\rho^2 \|\mathbf{h^{\dagger}}_{RA} \mathbf{w}_t\|^2 + \|\mathbf{h^{\dagger}}_{RA} \mathbf{w}_t\|^2 + P_A| {h_{AA}}|^2 + 1}\right)\notag\\ 
&~~~~~~~~~~~~~~~~~~~~~~
  +  \log_2 \left( 1 +  \frac{\rho^2 P_A C_{rA} |\mathbf{h^{\dagger}}_{RB} \mathbf{w}_t|^2 }{\rho^2 \|\mathbf{h^{\dagger}}_{RB} \mathbf{w}_t\|^2 + \|\mathbf{h^{\dagger}}_{RB} \mathbf{w}_t\|^2 + P_B| {h_{BB}}|^2 + 1}\right)\\ 
   {\rm s.t.} &~~
  (1-\rho)(|\mathbf{h}_{AR}|^2{P_A} + |\mathbf{h}_{BR}|^2{P_B}+ \mathrm{\bar{E}} + 1) \geq \bar{Q},\\
 &~~ \rho (P_A \|\mathbf{w}_t\|^2C_{rA} + P_B \|\mathbf{w}_t\|^2C_{rB} + \|\mathbf{ w}_t\|^2 ) +  \|\mathbf {w}_t\|^2 \leq P_R,\\
&~~  \mathbf{w}_r^{\dagger}\mathbf{H}_{RR}\mathbf{w}_t=0,
\end{align}
\end{subequations}
\hrulefill 
\end{figure*}
where $C_{rA}\triangleq |\mathbf{w}_r^{\dagger}\mathbf{h}_{AR}|^2 $ and $C_{rB}\triangleq |\mathbf{w}_r^{\dagger} \mathbf{h}_{BR}|^2.$ 
 
\subsection{Parameterization of Receive Beamforming}
Observe in  (\ref{y6}) that $\mathbf{w}_r$ is mainly involved in $|\mathbf{w}_r^{\dagger}\mathbf{h}_{AR}|^2$ and $|\mathbf{w}_r^{\dagger}\mathbf{h}_{BR}|^2$, so it has to balance the signals received from the sources. According to the result obtained in \cite{complete}, $\mathbf{w}_r$ can be parameterized by $0 \leq \alpha \leq 1$ as 
\begin{equation}
\mathbf{w}_r = \alpha \frac{\Pi_{\mathbf{h}_{BR}}\mathbf{h}_{AR}}{\|\Pi_{\mathbf{h}_{BR}}\mathbf{h}_{AR}\|} + \sqrt{1-\alpha}\frac{\Pi^{\perp}_{\mathbf{h}_{BR}}\mathbf{h}_{AR}}{\|\Pi^{\perp}_{\mathbf{h}_{BR}}\mathbf{h}_{AR}\|},\label{alex1}
\end{equation}
where $\alpha$ is a non-negative real-valued scaler.

It should be made clear that (\ref{alex1}) is not a complete characterization of $\mathbf{w}_r$ because it is also involved in the ZF constraint $\mathbf{w}_r^{\dagger}\mathbf{H}_{RR}\mathbf{w}_t= 0,$ but this parameterization makes the problem more tractable. Thus, given $\alpha,$ we can optimize $\mathbf{w}_t$ for fixed PS ratio $\rho$. Then perform a 1-D search to find the optimal $\alpha^*$. 

 \subsection{Optimization of the Receive PS Ratio}
For given $\mathbf{w}_r$ and $\mathbf{w}_t$, the optimal receive PS ratio $\rho$ can be determined. Firstly, using the monotonicity between SINR and the rate, (\ref{y6}) can be rewritten as
\begin{align}
 \max_{{ \rho \in(0,1)}}  &~~  \frac{\rho^2 P_B C_{rB}|\mathbf{h^{\dagger}}_{RA} \mathbf{w}_t|^2 }{\rho^2 \|\mathbf{h^{\dagger}}_{RA} \mathbf{w}_t\|^2 + \|\mathbf{h^{\dagger}}_{RA} \mathbf{w}_t\|^2 + P_A| {h}_{AA}|^2 + 1}+\notag\\ 
&~~ \frac{\rho^2 P_A C_{rA} |\mathbf{h^{\dagger}}_{RB} \mathbf{w_t}|^2 }{\rho^2 \|\mathbf{h^{\dagger}}_{RB} \mathbf{w}_t\|^2 + \|\mathbf{h^{\dagger}}_{RB} \mathbf{w}_t\|^2 + P_B| {h_{BB}}|^2 + 1}\notag\\
{\rm s.t.} &~~(1-\rho)(|\mathbf{h}_{AR}|^2{P_A} + |\mathbf{h}_{BR}|^2{P_B}+ \mathrm{\bar{E}} + 1) \geq \bar{Q},\notag\\
 &~~ \rho (P_A \|\mathbf{w}_t\|^2C_{rA} + P_B \|\mathbf{w}_t\|^2C_{rB} + \|\mathbf{w}_t\|^2 )\notag\\
 &~~~~~~~~~~~~~~~~~~~~~~~~ +  \|\mathbf {w}_t\|^2 \leq P_R. \label{rho1}
\end{align}
Problem (\ref{rho1}) is a linear-fractional programming problem, and can be converted into a linear programming problem \cite{Boyd}. The receive PS ratio is determined by the equation set below:
\begin{subequations}
\begin{eqnarray}
 (1-\rho)(|\mathbf{h}_{AR}|^2{P_A} + |\mathbf{h}_{BR}|^2{P_B}+ \mathrm{\bar{E}} + 1) \geq \bar{Q},\\
\rho (P_A \|\mathbf{w}_t\|^2C_{rA} + P_B \|\mathbf{w}_t\|^2C_{rB} + \|\mathbf{w}_t\|^2 ) +  \|\mathbf {w}_t\|^2\nonumber\\
\leq P_R  \label{lola}
\end{eqnarray}
\end{subequations}
Using the procedure in  \cite{joint}, the optimal $\rho$ can be found by
\begin{equation}
\rho^* \leq \frac{P_R - \|\mathbf{w}_t\|^2 }{P_A \|\mathbf{w}_t\|^2C_{rA} + P_B \|\mathbf{w}_t\|^2C_{rB} + \|\mathbf{w}_t\|^2}.
\end{equation}
We check whether the above solution satisfies the constraint (\ref{rho1}). If it does, then it is the optimal solution. Otherwise the energy harvesting constraint should be met with equality thus giving the optimal receive PS ratio $\rho^*$ given by 
\begin{equation}
 \rho^* = 1 - \frac{\bar{Q}}{|\mathbf{h}_{AR}|^2{P_A} + |\mathbf{h}_{BR}|^2{P_B}+ \mathrm{\bar{E}} + 1}.
\end{equation}

\subsection{Optimization of Transmit Beamforming}
Here, we first study how to optimize $\mathbf{w}_t$ for given $\alpha$ and $\rho.$  Then we perform a 1-D search on $\alpha$ to find the optimal $\alpha^*$ which guarantees an optimal $\mathbf{w}_r^*$  as defined in (\ref{alex1}) for the given $\rho.$
For convenience, we define a semidefinite matrix $\mathbf{W}_t= \mathbf{w}_t\mathbf{w}_t^{\dagger}$. Then problem (\ref{y6}) becomes
\begin{eqnarray}
 \max_{{\mathbf{W}_t\succeq0}} \!\!\!& &\!\!\! F(\mathbf{W}_t) \nonumber\\
  {\rm s.t.}  
  \!\!\!& &\!\!\! \mathrm{trace} (\mathbf{W}_t) \leq \frac{P_R}{\rho(P_A C_{rA}+ P_B C_{rB}+1)+1}\nonumber\\
   \!\!\!& &\!\!\! (1-\rho)(|\mathbf{h}_{AR}|^2P_A + |\mathbf{h}_{BR}|^2P_B + \mathrm{\bar{E}} + 1)\nonumber\\
 \!\!\!& &\!\!\!  \geq \bar{Q}\nonumber\\
 \!\!\!& &\!\!\!   \mathrm{trace} (\mathbf{W}_t\mathbf{H}_{RR}^{\dagger} \mathbf{w}_r\mathbf{w}_r^{\dagger}\mathbf{H}_{RR}) \nonumber\\ 
\!\!\!& &\!\!\!  \mathrm{rank}(\mathbf{W}_t) = 1,\label{max1}  
\end{eqnarray}
where  $F(\mathbf{W}_t)$ is given in (\ref{FW}) (see next page).
 \begin{figure*}[!t]
\normalsize
\begin{multline}\label{FW}
F(\mathbf{W}_t)\triangleq 
 \log_2\left( 1+ \frac{\rho^2 P_B C_{rB}\mathrm{trace}(\mathbf{W}_t\mathbf{h}_{RA} \mathbf{h}_{RA}^{\dagger}) }{\rho^2 \mathrm{trace}(\mathbf{W}_t \mathbf{h}_{RA} \mathbf{h}_{RA}^{\dagger}) + \mathrm{trace}( \mathbf{W}_t\mathbf{h}_{RA} \mathbf{h}_{RA}^{\dagger}) + P_A| {h_{AA}}|^2 + 1}\right)\\
 +\log_2 \left( 1 +  \frac{\rho^2 P_A C_{rA} \mathrm{trace}( \mathbf{W}_t\mathbf{h}_{RB}   \mathbf{h}_{RB}^{\dagger}) }{\rho^2 \mathrm{trace}(\mathbf{W}_t \mathbf{h}_{RB} \mathbf{h}_{RB}^{\dagger}) + \mathrm{trace} (\mathbf{W}_t \mathbf{h}_{RB}\mathbf{h}_{RB}^{\dagger} ) + P_B| {h_{BB}}|^2 + 1}\right)
 \end{multline}
\hrulefill 
\end{figure*}
Clearly, $F(\mathbf{W}_t)$ is not a concave function, making the problem challenging. To solve (\ref{FW}), we propose to use the difference of convex programming (DC) to find a local optimum point. To this end, we express $F(\mathbf{W}_t)$ as a difference of two concave functions $f(\mathbf{W}_t)$ and $g(\mathbf{W}_t)$ \cite{joint}, i.e.,
\begin{multline}
 F(\mathbf{W}_t)=
 \log_2((\rho^2P_B C_{rB}+ \rho^2 +1)\mathrm{trace}(\mathbf{W}_t\mathbf{h}_{RA}\mathbf{h}_{RA}^{\dagger})\\
+P_A |{h_{AA}}|^2 + 1)-\log_2(\rho^2 \mathrm{trace}(\mathbf{W}_t\mathbf{h}_{RA}\mathbf{h}_{RA}^{\dagger})\\
+\mathrm{trace}(\mathbf{W}_t\mathbf{h}_{RA}\mathbf{h}_{RA}^{\dagger})+ P_A |{h_{AA}}|^2 + 1)\\
+\log_2((\rho^2P_A C_{rA}+ \rho^2 +1)\mathrm{trace}(\mathbf{W_t}\mathbf{h}_{RB}\mathbf{h}_{RB}^{\dagger})\\
+P_B |{h_{BB}}|^2 + 1)- \log_2(\rho^2 \mathrm{trace}(\mathbf{W}_t\mathbf{h}_{RB}\mathbf{h}_{RB}^{\dagger})\\
+\mathrm{trace}(\mathbf{W}_t\mathbf{h}_{RB}\mathbf{h}_{RB}^{\dagger})+P_B |{h_{BB}}|^2 + 1)\\
\triangleq f(\mathbf{W}_t)- g(\mathbf{W}_t),
\end{multline}   
where $f(\mathbf{W}_t) \triangleq \log_2((\rho^2P_B C_{rB}+ \rho^2 +1)\times$$\mathrm{trace}(\mathbf{W}_t\mathbf{h}_{RA}\mathbf{h}_{RA}^{\dagger})$$+ P_A |{h_{AA}}|^2$$ + 1)+ \log_2((\rho^2P_A C_{rA}+ \rho^2 +1)\mathrm{trace}(\mathbf{W}_t\mathbf{h}_{RB}\mathbf{h}_{RB}^{\dagger})$$+P_B |{h_{BB}}|^2 + 1)$ and $g(\mathbf{W}_t)\triangleq  \log_2(\rho^2 \mathrm{trace}(\mathbf{W}_t\mathbf{h}_{RA}\mathbf{h}_{RA}^{\dagger})$$+\mathrm{trace}(\mathbf{W}_t\mathbf{h}_{RA}\mathbf{h}_{RA}^{\dagger})$$+  P_A |{h_{AA}}|^2 + 1)$
$+ \log_2(\rho^2 \mathrm{trace}(\mathbf{W}_t\mathbf{h}_{RB}\mathbf{h}_{RB}^{\dagger})$
$+ \mathrm{trace}(\mathbf{W}_t\mathbf{h}_{RB}\mathbf{h}_{RB}^{\dagger})+ P_B |{h_{BB}}|^2 + 1)$. Note that $f(\mathbf{W}_t)$ is a concave function while $g(\mathbf{W}_t)$ is a convex function. The main idea is to approximate $g(\mathbf{W}_t)$ by a linear function. The linearization (first-order approximation) of $g(\mathbf{W}_t)$ around the point $f(\mathbf{W}_{t,k})$ is given in (\ref{main^_prob}).
\begin{figure*}[!t]
\normalsize
\begin{multline}\label{main^_prob}
g_L(\mathbf{W}_t; \mathbf{W}_{t,k}) = 
 \frac{1}{\mathrm{In}(2)} \frac{\rho^2\mathrm{trace}{((\mathbf{W}_t-\mathbf{W}_{t,k})\mathbf{h}_{RA}\mathbf{h}_{RA}^{\dagger})}+ \mathrm{trace}((\mathbf{W}_t-\mathbf{W}_{t,k})\mathbf{h}_{RA}\mathbf{h}_{RA}^{\dagger})}{\rho^2 \mathrm{trace}(\mathbf{W}_{t,k}\mathbf{h}_{RA}\mathbf{h}_{RA}^{\dagger})+ \mathrm{trace}(\mathbf{W}_t\mathbf{h}_{RA}\mathbf{h}_{RA}^{\dagger})+ P_A|{h_{AA}}|^2+1}\nonumber\\
 +\frac{1}{\mathrm{In}(2)} \frac{\rho^2\mathrm{trace}{((\mathbf{W}_t-\mathbf{W}_{t,k})\mathbf{h}_{RB}\mathbf{h}_{RB}^{\dagger})}+ \mathrm{trace}((\mathbf{W}_t-\mathbf{W}_{t,k})\mathbf{h}_{RB}\mathbf{h}_{RB}^{\dagger})}{\rho^2 \mathrm{trace}(\mathbf{W}_{t,k}\mathbf{h}_{RB}\mathbf{h}_{RB}^{\dagger})+ \mathrm{trace}(\mathbf{W}_t\mathbf{h}_{RB}\mathbf{h}_{RB}^{\dagger})+ P_B|{h_{BB}}|^2+1}\nonumber\\
 +\log_2(\rho^2 \mathrm{trace}(\mathbf{W}_{t,k}\mathbf{h}_{RA}\mathbf{h}_{RA}^{\dagger}) + \mathrm{trace}(\mathbf{W}_{t,k}\mathbf{h}_{RA}\mathbf{h}_{RA}^{\dagger})+ P_A |{h_{AA}}|^2+1)\nonumber\\ 
 +\log_2(\rho^2 \mathrm{trace}(\mathbf{W}_{t,k}\mathbf{h}_{RB}\mathbf{h}_{RB}^{\dagger}) + \mathrm{trace}(\mathbf{W}_{t,k}\mathbf{h}_{RB}\mathbf{h}_{RB}^{\dagger})+ P_B |{h_{BB}}|^2+1).
\end{multline}
\hrulefill 
\end{figure*}
Then the DC programming is applied to sequentially solve the following convex problem:
\begin{eqnarray}
\mathbf{W}_{t,k+1} \!\!\!&=&\!\!\! \mbox{arg} \max_{{\mathbf{W_t}}} f(\mathbf{W}_t) - g_L (\mathbf{W}_t; \mathbf{W}_{t,k}) \nonumber\\ {\rm s.t.} 
\!\!\!& &\!\!\! \mathrm{trace}(\mathbf{W}_t) = \frac{P_R}{\rho(P_A C_{rA}+ P_B C_{rB}+ 1)+1} \nonumber\\
 \!\!\!& &\!\!\!(1-\rho)(|\mathbf {h}_{AR}|^2P_A + |\mathbf{h}_{BR}|^2P_B+ \mathrm{\bar{E}} + 1)  \geq \bar{Q}  \nonumber\\
  \!\!\!& &\!\!\! \mathrm{trace}(\mathbf{W}_t\mathbf{H^{\dagger}}_{RR}\mathbf{w}_r\mathbf{w^{\dagger}}_r\mathbf{H}_{RR})=0.\label{main_prob2}
\end{eqnarray} 
We solve (\ref{main_prob2}) by:
\begin{itemize}
\item[(i)] Choosing an initial point $\mathbf{W}_t$.
\item[(ii)] For $K = 0, 1,\dots$, solve (\ref{main_prob2}) until convergence. Notice that in (\ref{main_prob2}),  we have ignored the rank--1 constraint on $\mathbf{W}_t.$ This constraint is guaranteed to be satisfied by the results in Theorem 2 in \cite {new_result} and also in \cite{ruhuul, ruhul_inter} when M$_T > 2$. Thus, the decomposition of $\mathbf{W}$ leads to the optimal solution $\mathbf{w}_t^{\dagger}$. 
\end{itemize}

\subsection{Optimization of Receive Beamforming}
Given $\mathbf{w}_t$, the optimal receive beamforming $\mathbf{w}_r$ can be obtained by performing a 1-D search on $\alpha$ to find the maximum $\alpha^*$ which maximizes $R_{sum}(\mathbf{w}_r)$ for a fixed value of $\rho\in (0,1)$. See Algorithm 1. The bounds of the rate search interval are obtained as follows. The lower bound $(R_A + R_B)_{low}$ is obviously zero while the upper bound $(R_A + R_B)_{max}$ is defined as the achievable sum-rate at zero RSI. With optimal $\alpha^*$, the optimal $\mathbf{w}^*_r$ can be obtained from (\ref{alex1}).

\subsection{Iterative Update}
Now, the original beamforming and receive PS optimization in problem (\ref{y6}) can be solved by an iterative technique shown in Algorithm 2. Algorithm 2 continually updates the objective function in (\ref{y6}) until convergence.

\begin{algorithm}[ht]
\caption{Procedure for Solving (\ref{max1})}
\label{algorithm1}
\begin{algorithmic}[1]
\State Set  $\bar{Q}> 0,$
 $\delta^2_{p}=1.$ Set $R_{sum}= R_A + R_B,$ as numerals and $R_{\rm diff}=$ (any value $> R_{sum}$). Set $k = 0.$
 \State Obtain $\rho^*$ by considering (\ref{lola}). 
 \State Set $\alpha$ = non-negative scaler and obtain $\mathbf{w}_r$ in (\ref{alex1}).
\State At step $k$, set $\alpha(k) = \alpha(k - 1) + \triangle\alpha$ until $\alpha(k)=1,$ where $\triangle \alpha$ is the searching step size.
 \State Set $(R_A + R_B)_{low} = 0$, $(R_A + R_B)_{up} = (R_A + R_B)_{max}$
 \State \textbf{Repeat}
 
 		 i) Set sum-rate $\leftarrow \frac{1}{2}((R_A + R_B)_{low} +  (R_A + R_B)_{up})$
 		 
 		ii) Obtain the optimal $\mathbf{W}_t$ by solving (\ref{main_prob2}).
 		 
		iii) Update the value of $(R_A + R_B)$ with the bisection 
		
		search method: if (ii) is feasible, set $(R_A + R_B)_{low} =$ 
		
		$R_{sum}$; otherwise, $(R_A + R_B)_{up} = R_{sum}$.
 \State \textbf{Until}
 $(R_A + R_B)_{up} - (R_A + R_B)_{low} < \epsilon,$ where $\epsilon$ is a
 
  small positive number. Thus we get the optimal $\alpha$
  
   which maximizes (\ref{alex1}) to give $R_{sum}(\mathbf{w}_r(k)).$
	   
\State $k = k+1$
\State Obtain  $R_{sum}(\mathbf{w}_r^*)$ by comparing $R_{sum}(\alpha(k)),~\forall k$.
\end{algorithmic}
\end{algorithm}

\begin{algorithm}[ht]
\caption{Procedure for Solving (\ref{y6})}
\label{alg2}
\begin{algorithmic}[1]
\State Initialise  $\alpha$
\State \textbf{Repeat}

		1) Solve (\ref{rho1}) to obtain optimal $\rho.$
		
		2) Solve (\ref{max1}) using Algorithm \ref{algorithm1} to obtain
		
		 $\mathbf{w}_t^*,$ $\mathbf{w}_r^*,$ $R_{sum}(\mathbf{w}_r^*).$
\end{algorithmic}
\end{algorithm}

\section{Numerical example}\label{num_1}
\begin{figure}[ht]
\centering
\includegraphics*[width=5cm]{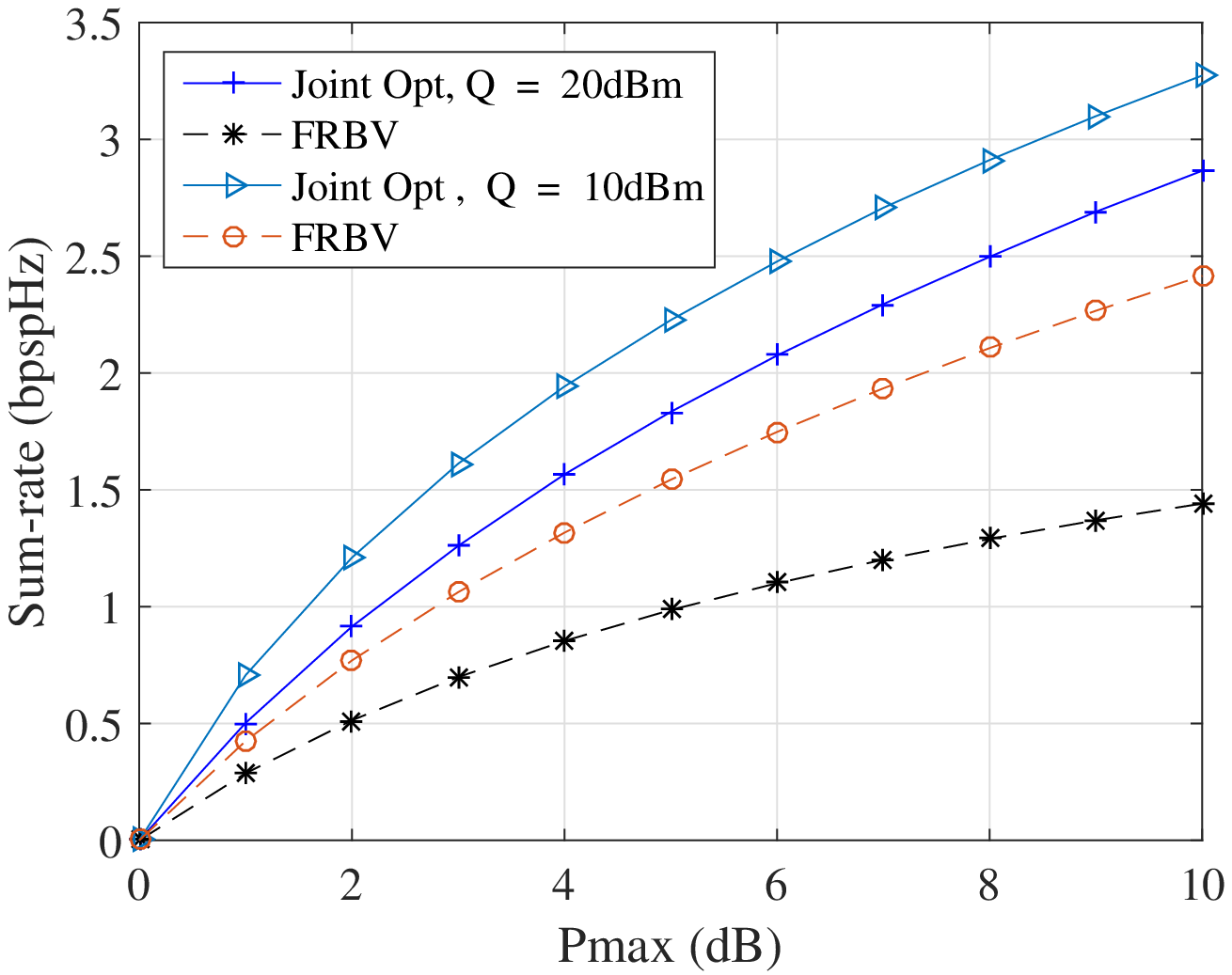}
\caption{Sum-rate versus P$_{\mbox{max}}$.}
\label{SumR_eupisco}
\end{figure}

\begin{figure}[ht]
\centering
\includegraphics*[width=5cm]{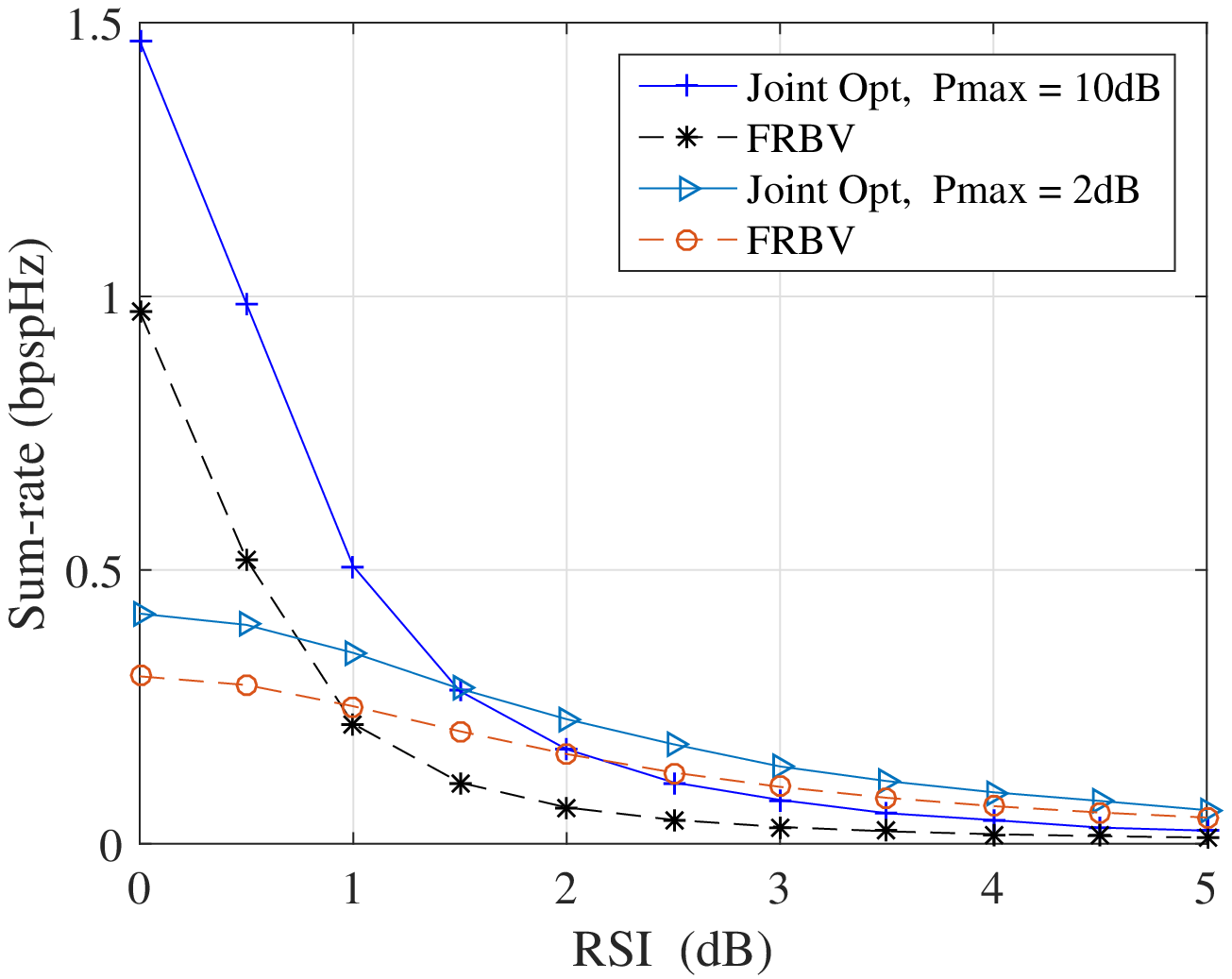}
\caption{Sum-rate versus RSI.}
\label{SI_eupisco}
\end{figure}

In this section, we evaluate the performance of the proposed algorithm through computer simulations assuming flat Rayleigh fading environments. In Fig.~\ref{SumR_eupisco}, we show the sum-rate results against versus the transmit power budget $P_{\mbox{max}}$ (dB) for various harvested energy constraint. The proposed scheme (`Joint Opt' in the figure) is compared with those of the fixed receive beamforming vector ($\mathbf{w}_r$) (`FRBV'= 0.583) at optimal PS coefficient ($\rho^*$). Remarkably, the proposed scheme yields higher sum-rate compared to the sum-rate of the FRBV schemes which essentially necessitates joint optimization. The impact of the RSI on the sum-rate is studied in Fig.~\ref{SI_eupisco}. Results show that an increase in the RSI results in a corresponding decrease in the achievable sum-rate.

\section{Conclusion}\label{conc_1}
In this paper, we investigated the joint beamforming optimization for SWIPT in FD MIMO two-way relay channel and proposed an algorithm which maximizes the sum-rate subject to the relay transmit power and harvested energy constraints. Using DC and a 1-D search, we jointly optimized the receive beamforming vector, the transmit beamforming vector, and receive PS ratio to maximize the sum-rate. Simulation results confirm the importance of joint optimization.

\end{document}